# Carrier density independent scattering rate in SrTiO$_3$-based electron liquids

Evgeny Mikheev[1], Santosh Raghavan[1], Jack Y. Zhang[1], Patrick B. Marshall[1], Adam P. Kajdos[1], Leon Balents[2], and Susanne Stemmer[1]

[1]Materials Department, University of California, Santa Barbara, CA 93106-5050, USA

[2]Kavli Institute for Theoretical Physics, University of California, Santa Barbara, California 93106-4030, USA

**Abstract**

We examine the carrier density dependence of the scattering rate in two- and three-dimensional electron liquids in SrTiO$_3$ in the regime where it scales with $T^n$ ($T$ is the temperature and $n \leq 2$) in the cases when it is varied by electrostatic control and chemical doping, respectively. It is shown that the scattering rate is independent of the carrier density. This is contrary to the expectations from Landau Fermi liquid theory, where the scattering rate scales inversely with the Fermi energy ($E_F$). We discuss that the behavior is very similar to systems traditionally identified as non-Fermi liquids ($n < 2$). This includes the cuprates and other transition metal oxide perovskites, where strikingly similar density-independent scattering rates have been observed. The results indicate that the applicability of Fermi liquid theory should be questioned for a much broader range of correlated materials and point to the need for a unified theory.

## I. Introduction

A complete understanding of materials whose transport and thermodynamic properties deviate from Landau Fermi liquid theory remains one of the central problems in physics [1,2]. Non-Fermi-liquid behavior is usually identified via power-laws in the temperature ($T$) dependence of properties such as electrical resistance. For example, the resistance of non-Fermi liquids scales with $T^n$, where $n < 2$. In many cases, including some very extensively characterized materials, such as the cuprate superconductors, it is still under debate if the observed behavior can be described by a modified Fermi liquid theory or if a completely new description of such metallic states is required [1-4].

Conversely, a correlated electron system is usually identified as a classic Fermi liquid when $n = 2$, which is the exponent predicted from the electron-electron scattering contribution to the resistance. There may, however, exist another case, namely systems that *appear* to behave like a Fermi liquid, exhibiting $n = 2$, but on closer inspection fail to be described by Fermi liquid theory. This case is important, because many of the systems that exhibit non-Fermi liquid behavior and exotic states, such as unconventional superconductivity, emerge from "parent phases" that are thought to be Fermi liquids, because $n = 2$.

The fact that the resistance follows $T^2$ behavior up to unusually high temperatures ($T \gg T_F$, where $T_F$ is the Fermi energy) in many correlated electron systems may already hint at difficulties for Fermi liquid theory. Examples include highly doped $SrTiO_3$ [5-8], cuprate superconductors that show $T^2$ behavior to room temperature in some parts of their phase diagrams [9,10], and rare earth nickelates, which follow $n = 2$ to room temperature [11,12]. Well-defined $T^2$ resistivity is found in many other 3$d$, 4$d$, and 5$d$ transition metal perovskites [13-15]. Another phenomenon that is difficult to reconcile with Fermi liquid quasiparticles [16-18] is a pronounced separation of the 0-K Hall and longitudinal transport scattering rates in electron liquids in SrTiO$_3$ quantum wells even though $n \approx 2$ indicates a Fermi liquid [19].

In this work, we examine the applicability of Fermi liquid theory by tuning the carrier density in two different SrTiO$_3$-based electron liquids: (i) *chemically doped* SrTiO$_3$ and (ii) electrostatically gated two-dimensional electron liquids (2DELs) in thin SrTiO$_3$ *quantum wells* sandwiched between two insulating SmTiO$_3$ layers. Uniformly doped SrTiO$_3$ exhibits a $T^2$ dependence of the resistance over an exceptionally wide doping range [5-8,20]. As the doping is increased, optical phonon mode scattering, which operates at high temperatures, is screened out and the $T^2$ regime expands to ever higher temperatures [7]. This behavior is consistent with electron-electron scattering dominating the transport. The quantum wells contain about $7\times10^{14}$ cm$^{-2}$ mobile carriers that are introduced by the charge (polar) discontinuity between SrTiO$_3$ and SmTiO$_3$ [21]. Electrical transport is dominated by the $T^n$ contribution ($n \leq 2$) to room temperature [7,22]. A drop from $n \sim 2$ to $n \sim 5/3$ occurs at a critical quantum well thickness [19]. The endpoint (SmTiO$_3$ without any embedded SrO layer) is an antiferromagnetic Mott insulator.

We examine the temperature coefficient $A$, the residual resistance $R_0$, and the exponent $n$ in the temperature dependence of the electrical sheet resistance, $R_{xx}$, as the 2D or 3D carrier density is varied:

$$R_{xx} = R_0 + AT^n \tag{1}$$

We use electric field gating to modulate the charge carrier densities in the 2DELs. This has the advantage of avoiding alloy disorder and allows for continuous modulation of the carrier density by the applied electric voltage. We show that Fermi liquid theory faces difficulties in providing a quantitative explanation of a nearly carrier density independent scattering rate in $R_{xx}$ for both the 2DELs and the uniformly doped SrTiO$_3$.

## Results

Figures 1(a) shows a schematic of the gated Hall bar device structure. An optical micrograph is shown in Fig. 1(b). A 6-nm layer of the wide-band gap insulator $SrZrO_3$ served as a gate dielectric. The thin $SrTiO_3$ quantum well, containing one unit cell of $SrTiO_3$ (2 SrO layers), is confined by insulating $SmTiO_3$ layers. At this thickness of the $SrTiO_3$ quantum wells a non-Fermi liquid exponent $n$ is expected [19,23].

Figure 1(c) shows the temperature dependence of the quantum well sheet resistance, $R_{xx}$, at different gate voltages, $V_G$. The normalized $[(R/R(V_G = 0)]$ sheet resistance is shown in Fig. 1(d) across the entire measured $T$ range (2-300 K). The resistance modulation between $V_G = -1$ and $+1$ V was 3 - 4% and was reversible. The gate leakage current, $I_G(V_G)$, showed non-linear behavior at all temperatures [Fig. 1(e)]. The highest leakage was 15 nA, below 1 nA for most $V_G$'s, and in all cases orders of magnitude below the source-drain current of 40 µA.

The electrostatic gate effect was independently confirmed by Hall effect measurements as a function of $V_G$. The Hall resistance was linear in magnetic field, allowing for extraction of the carrier density $N$, which is shown in Fig. 1(f). The changes in $N$ were consistent with accumulation (depletion) of charge carriers for positive (negative) $V_G$, respectively, and the observed resistance change. The total modulation between $V_G = -1$ and $+1$ V was $\Delta N = 1.62 \times 10^{13}$ cm$^{-2}$. Despite this being a fairly large amount of charge modulated, due to the very high carrier density (~$3 \times 10^{14}$ cm$^{-2}$ per interface, corresponding to the theoretically expected ½ of an electron per interface unit cell [21]), the fractional modulation is only 2.1%. This was, however, sufficient to investigate the changes in the non-Fermi liquid parameters.

## Discussion

Figure 2 shows an example of a fit of $R_{xx}(T)$ at $V_G = 0$ V with $n = 1.645$. The results for all $V_G$ are shown in Fig. 3. The fit range was restricted to 150 - 240 K, but the description was valid in the 125 - 300 K range. Below 125 K, $R_{xx}(T)$ exhibited a logarithmic upturn, which is not yet understood. When $R_0$, $A$, and $n$ were used as free parameters, $n$ was insensitive to $V_G$ and thus the carrier density modulation, within experimental error [see error bars in Fig. 3(a)]. We note that these results are fully consistent with a much larger, systematic study of resistivity and Hall effect in ungated $RTiO_3/SrTiO_3/RTiO_3$ ($R$ = Sm, Gd) quantum wells as a function of $SrTiO_3$ thickness presented elsewhere [19].

The non-Fermi liquid exponent $n \sim 5/3$ has been observed in several other systems, such as electron-doped cuprates [24] and rare earth nickelates [11,12]. All systems are in proximity to an antiferromagnetically ordered state. Nevertheless, $n$ is at variance with theoretical predictions of two and three-dimensional antiferromagnetic quantum critical fluctuations ($n$ = 1 and 3/2, respectively [25]). In principle, $n$ could be modified from a theoretical value by disorder [26]. However, while disorder (interface roughness) is clearly reflected in $R_0(V_G)$ (see below), it does not measurably change the non-Fermi liquid exponent $n$. This points to an intrinsic origin of this particular value of $n$.

For reliable analysis of the trends in the $A$ parameter, $n$ was fixed to 1.645. As shown in Figs. 3(b) and (c), both $R_0$ and $A$ depend on $V_G$ and therefore $N$, as expected. We discuss $R_0(V_G)$ first. Specifically, $R_0 = m^*\Gamma_0/Ne^2$, where $m^*$ is the effective electron mass, $\Gamma_0$ is the scattering rate at $T = 0$ (a measure of disorder level) and $e$ the electron charge [27]. Figure 3(d) shows $R_0(V_G)$, $A(V_G)$, and $N^{-1}(V_G)$ normalized to their values at $V_G = 0$ V. $R_0(V_G)$ varies by a factor of two more than $N(V_G)$, implying that $\Gamma_0$ also changes with $V_G$. $\Gamma_0$ decreases upon charge carrier accumulation. This can be explained with asymmetries in interfacial roughness. Figure 4 shows a high-angle annular dark-field scanning transmission electron microscopy (HAADF-STEM) image, with interfaces indicated by the dashed lines. Overlaid on the image are averaged intensities of each column of atoms, integrated parallel to the interface. HAADF intensities are sensitive to atomic numbers present in the columns and thus provide a measure of intermixing/roughness. The intensities are relatively constant within the $SrTiO_3$ and $SmTiO_3$ layers. Interfacial regions on either side of the quantum well, marked by arrows, show a less abrupt intensity change for the bottom interface of the quantum well (right side in Fig. 5), indicating a rougher (more intermixed) interface. Asymmetries in top/bottom interface roughness are ubiquitous in superlattices [28,29]. In thin quantum wells, the dominant scattering contributing to $R_0$ is interface roughness [30]. For $V_G > 0$, charge carriers accumulate at the smoother top interface, thereby decreasing $\Gamma_0$ relative to $V_G < 0$.

We next discuss the dependence of $A$ in the carrier density $N$. Within the Drude model, the $AT^n$ term in Eq. (1) can be written as [31]:

$$AT^n = \frac{m^*}{Ne^2} \cdot \Gamma(T^n), \qquad (2)$$

where $\Gamma(T^n)$ is the scattering rate. From Figs. 3(d) and 5(a), one observes $A \sim N^{-1}$, in accordance with the pre-factor in Eq. (2). This implies that the scattering rate $\Gamma(T^n)$ is independent of $N$. This observation is at odds with the standard Fermi liquid theory of the electron-electron scattering contributions to the resistance, where the relevant energy scale is the Fermi energy, $E_F$. From dimensional analysis and $n = 2$ [31]:

$$\Gamma(T^2) = B \cdot \frac{(k_B T)^2}{\hbar E}, \qquad (3)$$

where $B$ is a dimensionless factor that incorporates scattering event probabilities and $k_B$ is the Boltzmann constant. For a Fermi liquid, $E = E_F$ [31]. For an arbitrary exponent ($1 \leq n \leq 2$), Eq. (3) becomes:

$$\Gamma(T^n) = B \cdot \frac{(k_B T)^n}{\hbar E^{n-1}}, \qquad (4)$$

In the case of the gated quantum well, $n = 1.645$ and $\Gamma(T^{1.645}) \sim 1/E^{0.645}$. Details of $B$ depend on the mechanism(s) giving rise to a momentum change and other assumptions [6,32-34]. In particular, Umklapp processes are necessary to relax the momentum [35] and their contribution is strongly dependent on $N$. For instance, according to refs. [33,36], $A \sim 1/N^{5/3}$. The importance of Umklapp processes has been discussed recently[8] for electron-electron scattering in $SrTiO_3$ doped to very low concentrations, where Umklapp processes are thought to be negligible (i.e., with much lower concentrations than the quantum wells studied here). There is debate in the literature as to whether a normal process can also relax momentum under certain conditions [37-39]. Independent of this question, however, from Luttinger's theorem, Eqs. (3-4) should result in a pronounced carrier density dependence of $\Gamma(T^n)$ *if* $E = E_F$, contrary to what is observed. The insensitivity of $\Gamma(T^n)$ to a change in $N$ upon gating therefore suggests that $E$ is not $E_F$, but some other, $N$ independent, energy scale.

To check if the scattering rate and the relevant energy scale $E$ remain insensitive to $N$ when it is varied over many orders of magnitude, we consider uniformly doped $SrTiO_3$. Figure 5(b) shows the $A$ as a function of $N$ (which is the 3D carrier density in this case) for different samples from the literature in the regime where $R_{xx}(T) \sim AT^2$. Again, we observe $A \sim N^{-1}$, implying that $\Gamma(T^2)$ is independent of $N$, at odds with Fermi liquid theory. The fact that this relationship is maintained over such a wide range of $N$ appears to also rule out any accidental cancellations by any carrier dependencies in $B$. We note weak steps in $A$, possibly associated with additional bands being filled [8], which was also observed in the temperature coefficient of the mobility [7], and could be due to an associated change in the density of states and/or multiband effect, both of which increase $A$. Nevertheless, the $A \sim N^{-1}$ relationship is maintained in each regime, even though the temperature exponent would suggest a classic Fermi liquid, i.e. $n = 2$.

In systems with $T$-linear behavior, such as the cuprates, $E$ in Eq. (4) becomes irrelevant as can be seen by setting $n = 1$. Furthermore, $B \sim 1$ [40,41]. This has sometimes been attributed to an underlying quantum critical point that causes the relaxation time to become independent of the microscopic processes so that the temperature is the only energy scale in the system [42]. The results here suggest something far more general, namely, a breakdown of the Fermi liquid state in certain materials, such as doped $SrTiO_3$, far from any (at least magnetic) quantum critical point. In these materials the energy scale in the scattering rate, which is not $E_F$, becomes increasingly irrelevant as a system transitions from an $n = 2$ regime to $n < 2$, as can be seen from Eq. (4).

It is remarkable that the "anomalous Fermi liquid" behavior is observed despite the fact that the materials studied here have well-defined Fermi surfaces [43-45] at low temperature, and show features consistent with filling of successive bands with increasing doping [43,45]. It is instructive to plot $\Gamma$, defined through Eq. (2), taking $m^* = m_e$, the free electron mass, as shown in Fig. 6. We see that it is always much smaller than the bandwidth (several eV). Furthermore, $\hbar\Gamma \gg k_B T$ at high temperature, which is clearly outside Fermi liquid theory. Moreover, if we estimate an order of magnitude for the energy scale by combining Eqs. 2 and 3, $E = (k_B T)^2/\hbar\Gamma$, we find $E \sim 10$ mV, which is smaller than $E_F$, at least at high carrier densities.

These observations compound the difficulty of conventional Fermi liquid theory in addressing the $T^2$ resistivity, and suggest that some distinctly stronger scattering process must be involved. Future studies should address this mechanism and the nature of the energy scale $E$ in Eqs. (3-4) in strongly correlated materials, such as the cuprates, nickelates, and titanates, all of which show a wide $T^2$ regime (as well as deviations from it). We note that in several materials with $n < 2$, which include the cuprates and the $SrTiO_3$ quantum wells, the Hall scattering rate maintains a $T^2$ dependence [18,19,46]. It has been pointed out that the nature of the energy scale ($E$) for the Hall scattering rate in such systems is unclear [17], but, interestingly, it has been found to have a weak doping dependence, i.e., similar to what we observe for $R_{xx}$ [9,17]. A doping independent scattering rate has also been recently reported in the cuprates [10] and several other perovskites [15]. Finally, we note that the continuity of the $A \sim 1/N$ behavior across four orders of magnitude of $N$ is inconsistent with a sudden, sharp increase in Umklapp scattering around $N \sim 2 \times 10^{20}$ cm$^{-3}$, as predicted in refs. 20 and 8.

In summary, the broad implication of this study is that some materials with a robust $T^2$ resistivity are not conventional Fermi liquids, and their transport is likely just as anomalous as that of those that are traditionally identified as non-Fermi liquids (when $1 \leq n \leq 2$). This leaves a formidable challenge to theory, namely a unified understanding of the entire $T^n$ resistivity regime, with a carrier density independent energy scale in the scattering rate, and allowing for intermediate exponents between the $T^2$ and the $T$-linear limits.

## Methods

The SmTiO$_3$(10 nm)/SrTiO$_3$(2 SrO layers thick)/SmTiO$_3$(1.2 nm)/SrZrO$_3$(6 nm) stack was grown on a (001)LSAT substrate by hybrid molecular beam epitaxy, as described in detail elsewhere [47-49]. The devices were processed using standard contact photolithography techniques. The Pt (100 nm) gate contact was deposited by electron beam evaporation. The mesa was defined by dry etching. Prior to the deposition of the Ti(400 nm)/Au(3000 nm) Ohmic contacts, SrZrO$_3$ was selectively removed with a wet etch in buffered HF diluted 1:10 in water. Electrical measurements from 2 to 300 K were performed in a Quantum Design Physical Property Measurement System (PPMS). Cross section samples for HAADF-STEM imaging were prepared by focused ion beam thinning and imaged using a FEI Titan S/TEM operated at 300 kV, with a convergence angle of 9.6 mrad. Atomic column positions were obtained by fitting each column to a two-dimensional Gaussian. Integrated intensities were obtained by averaging a circular region with radius ¼ of the unit cell around each atomic column [50].

## Acknowledgements

This work was supported in part by FAME, one of six centers of STARnet, a Semiconductor Research Corporation program sponsored by MARCO and DARPA and by the U.S. Army Research Office (grant no. W911NF-14-1-0379). L.B. and S.S. thank the MRSEC Program of the National Science Foundation under Award No. DMR-1121053, which supported the film growth experiments and the central facilities of the UCSB MRL used in this work. The microscopy studies (J. Y. Z.) were supported by the DOE (DEFG02-02ER45994). J.Y.Z. also received support from the Department of Defense through an NDSEG fellowship.

**Figure Legends**

**Figure 1:** (a) Schematic cross-sectional view of the test device. (b) Top view optical micrograph of the device. (c) Sheet resistance and (d) resistance normalized to $V_G = 0$ as a function of $T$ and $V_G$. (e) Gate leakage at different temperatures. (f) Carrier density measured by Hall effect at $T = 100$ K as a function of $V_G$.

**Figure 2:** (a) Non-Fermi liquid behavior in $R_{xx}(T)$ at $V_G = 0$. The dashed line is a fit to Eq. (1). (b) Same data and fit plotted as $dR_{xx}/dT$ vs. $T$. The inset is the same data on a log-log scale.

**Figure 3:** Results for the fit parameters in Eq. (1) as a function of $V_G$: (a) Non-Fermi liquid temperature exponent $n$, (b) residual resistance $R_0$ and, (c) the electron-electron scattering strength $A$. (d) $R_0$, $A$, and inverse carrier density $N^{-1}$ (at $T = 100$ K) normalized to their $V_G = 0$ values.

**Figure 4:** HAADF STEM image of a SrTiO$_3$ quantum well containing 4 SrO layers (orange dashed lines) sandwiched between two SmTiO$_3$ layers. The relative intensities of each plane of atoms, integrated parallel to the interface, are shown as an overlay. The arrows mark the first interfacial layer on either side of the quantum well. The more abrupt intensity difference between the SrO and SmO layers on the left side of the quantum well, compared to the right, indicates a sharper interface.

**Figure 5:** Temperature coefficient $A$ as a function of the carrier density $N$ in (a) gated quantum wells and (b) uniformly doped SrTiO$_3$. The dashed lines are a guide to the eye to illustrate $A \sim 1/N$ (blue) and $A \sim 1/N^{5/3}$ (grey). The latter relationship is often associated with electron-electron scattering in a Fermi liquid [33]. In (b), the carrier densities at which a higher lying conduction band fills, according to ref. [44], are indicated by vertical lines. A small increase in $A$ is observed at these carrier densities, but no deviation from $A \sim 1/N$. The experimental data in (b) are from La and Gd-doped $SrTiO_3$ thin films grown by molecular beam epitaxy [51,52] and from Nb-doped and oxygen deficient $SrTiO_3$ single crystals [6,8].

**Figure 6:** Scattering rate as a function of $T$, calculated as $\Gamma = ANT^2e^2/m^*$, for $m^* = m_e$, the free electron mass. $AN$ is taken from the slope of $A(N)$ data for uniformly doped $SrTiO_3$ in Fig. 5(b). The three lines correspond to the three regimes of band filling, as shown in Fig. 5(b). For comparison, the thermal energy, $k_BT$, is indicated by the black dashed line.

**Figure 1**

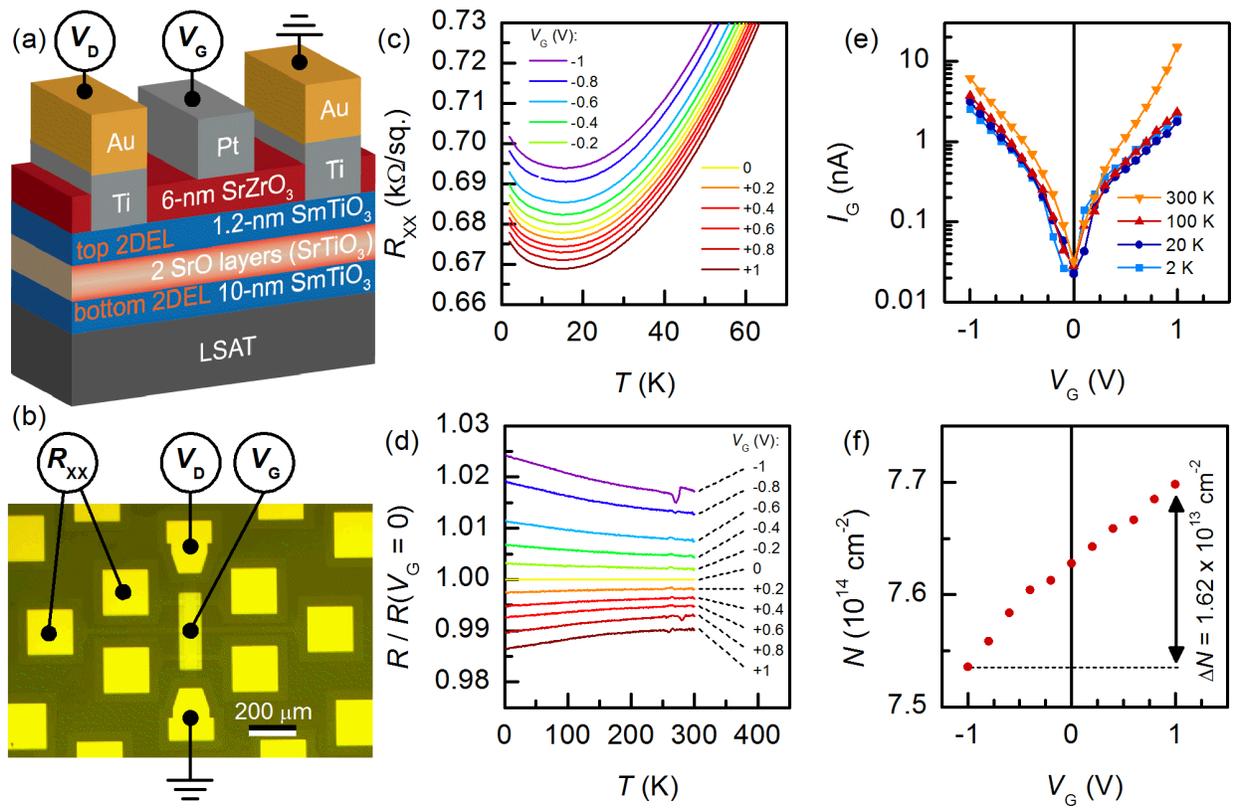

**Figure 2**

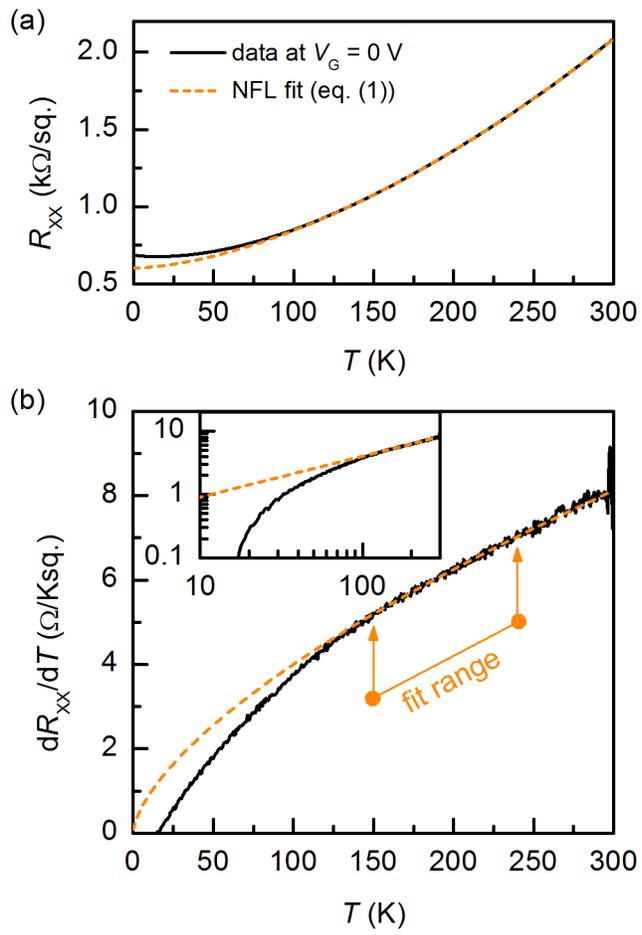

**Figure 3**

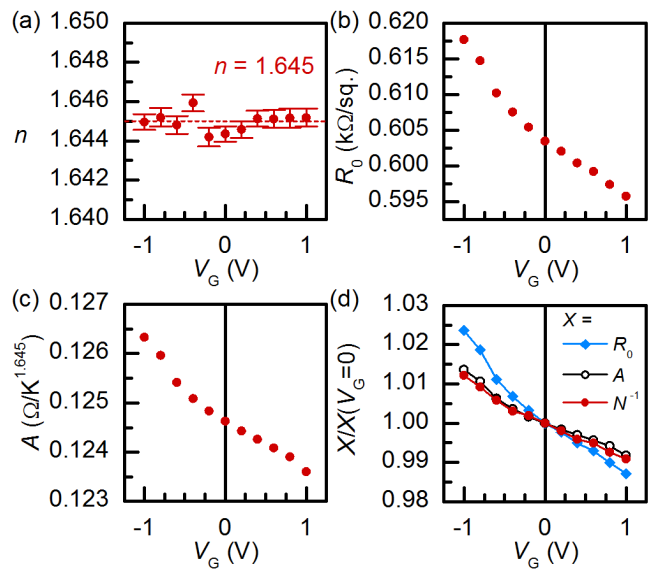

**Figure 4**

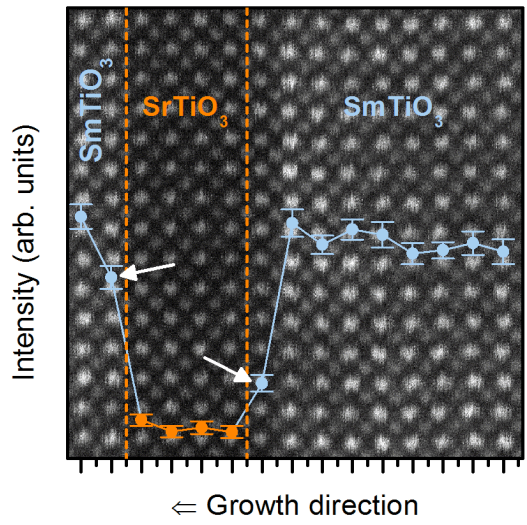

⇐ Growth direction

**Figure 5**

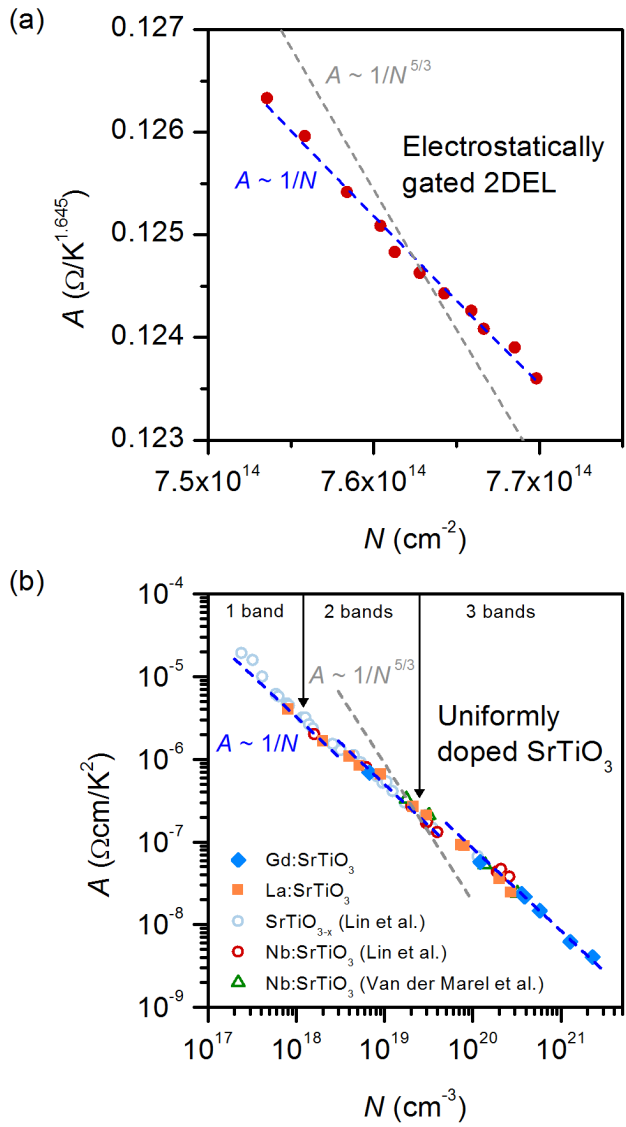

**Figure 6**

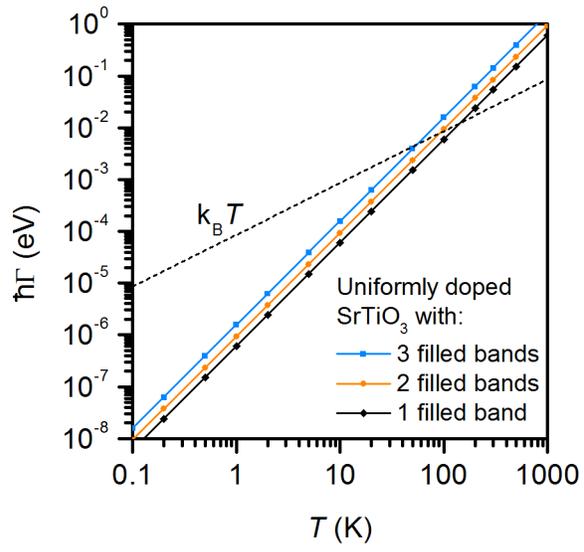